\documentclass[aps,prl,twocolumn,showpacs,groupedaddress]{revtex4}  % for review and submission 
\usepackage{graphicx}  % needed for figures
\usepackage{dcolumn}   % needed for some tables
\usepackage{bm}        % for math
\usepackage{amssymb}   % for math
\usepackage{amsmath}   % for math

\begin{document}

% The following information is for internal review, please remove them for submission
%\leftline{Version xx as of \today} 
%\leftline{Primary authors: Joe E. Physics}
%\leftline{To be submitted to (PRL, PRD-RC, PRD, PLB; choose one.)}
%\rightline{Comment to {\tt heremans@vt.edu}}

% the following line is for submission, including submission to the arXiv!!
%\hspace{5.2in} \mbox{Fermilab-Pub-04/xxx-E}

% Force line breaks with \\
%Lines break automatically or can be forced with \\

\title{Helical Aharonov-Casher edge states}

\author{J. J. Heremans}
\email{heremans@vt.edu}
\altaffiliation{Author to whom correspondence should be addressed: Department of Physics, Virginia Tech, Blacksburg, VA 24061, USA}
\author{L. L. Xu}
%\email{lx@vt.edu}
\affiliation{Department of Physics, Virginia Tech, Blacksburg, VA 24061, USA}

\date{\today}

\begin{abstract}
It is shown that an Aharonov-Casher vector potential in a two-dimensional geometry can lead to helical edge states. The Aharonov-Casher vector potential is the electromagnetic dual of the magnetic vector potential, and leads to traveling states at the sample edge in analogy to the integer quantum Hall effect. The helical edge states are predicted to appear in a narrow channel geometry with parabolic or sufficiently symmetric confinement potential. The implications of the helical Aharonov-Casher edge states and experimental considerations in specific materials systems are discussed. 
\end{abstract}

\maketitle 

The quantum Hall effects, both fractional and integer, occur in two-dimensional carrier systems (2DSs) upon application of a magnetic field $\bm{B}$ perpendicular to the carrier plane \cite{klitzingiqhe,tsuifqhe}. A description of the integer quantum Hall effect (IQHE) starts with the introduction of a magnetic vector potential, describing the interaction between a charge and a magnetic field.  We consider here the physical phenomena generated by replacing the magnetic vector potential by the Aharonov-Casher (AC) vector potential (defined below) describing the interaction between a magnetic moment or spin and an electric field. We find that in a narrow channel geometry one can then observe helical edge states akin to those characterizing the quantum spin Hall effect (QSHE) state \cite{kane2005b,bernevig2006b,koenig2007,fu2007,moore2007}. 

The IQHE can be described as arising from one-dimensional (1D) states at the sample edges, in which backscattering is forbidden \cite{halperin1982,macdonald1984}.  The edge states arise from a broken translational invariance induced by the edge, lifting the Landau level degeneracy and leading to propagating chiral edge states. The chirality implies that states propagating in opposite directions are found at opposite sample edges, with the spatial separation protecting the states from backscattering. The protection results in quantization of the Hall conductivity and vanishing longitudinal resistance. The QSHE also occurs at the edge of a 2DS, but its 1D edge states show a spin structure \cite{kane2005b,koenig2007}. The QSHE state and its three-dimensional topological insulator analogs \cite{kane2005b,bernevig2006b,koenig2007,fu2007,moore2007} have attracted considerable attention for the implications of their topologically protected states. Unlike in the IQHE where spin polarity does not affect edge state propagation direction, in the QSHE opposite spin polarities at the same edge propagate in opposite directions (helical edge states). The spin-polarized channels are protected from scattering unless time-reversal symmetry is broken, flipping the spin and leading to backscattering within one edge \cite{kane2005b}.  We will see that the AC vector potential generates such a spin structure in a narrow channel. A semi-classical context, briefly described here, provides insight into these phenomena. Classically, a Lorentz transformation can reduce the magnetic field creating the IQHE to zero as perceived by a moving observer, if also a electric field is present in the plane of the 2DS. Confinement of the carriers at the sample edge serves as the source of electric field. The magnetic vector potential yields the AC vector potential under Lorentz transformation, and concomitantly, the moving observer perceives the electrical charges as magnetic moments. The moving observer still notices edge phenomena, but the IQHE edge states are perceived as edge states of magnetic moments (although not necessarily helical). Spin is distinct from classical moments, and specific phenomena arise from spin, such as the helicity of states in the QSHE. We point out that, notwithstanding classical limitations, thought-experiments using the Lorentz transformation point to a mapping that may help in the description of other edge phenomena, since in principle the transformation converts between different types of physically non-trivial edge states. As example, beyond the scope of this paper, one may ask what a moving observer will conclude about the fractional quantum Hall effect \cite{herdminic2008}. 

\begin{figure}
\includegraphics[width=83mm]{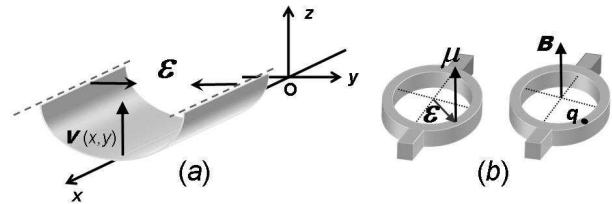}
\caption{\label{figluher11}(a): Sketch of the channel geometry, with parabolic potential at the edges. (b): The duality between the AB effect induced by $q\bm{A}$ and the AC effect induced by $(1/c^2)(\bm{\mu}\times\bm{\mathcal{E}})$, illustrated by interferometric ring geometries.}
\end{figure}

Referring to Fig.~\ref{figluher11}a, an IQHE experiment can be performed by applying $\bm{B}$ along the $z$-axis, perpendicular to a 2DS situated in the $x-y$ plane.  To derive IQHE edge states, a potential describing the sample edges must be introduced \cite{macdonald1984,zawadzki1989}. With the edge parallel to the $x$-axis, we introduce a parabolic confinement \cite{zawadzki1989,hattori2006,hsiao2009}, $V(x,y)=\frac{1}{2}m\,\omega_p^2\,y^2$. Hence a confinement electric field $\bm{\mathcal{E}}=(0,\mathcal{E},0)$ is present on both sides of the sample, with a $y$-component linear in $y$, $\mathcal{E}=-(m/q)\,\omega_p^2\,y$ (carrier charge and mass are represented by $q$ and $m$). The magnetic vector potential $\bm{A}$ enters the Hamiltonian in the term $(1/2m)(\bm{p}-q\bm{A})^2$. We now replace $q\bm{A}$ with its electromagnetic dual, the AC vector potential $\frac{1}{c^2}\bm{\mu}\times\bm{\mathcal{E}}$, describing the interaction of a magnetic moment $\bm{\mu}$ with electric field $\bm{\mathcal{E}}$. Here $c$ denotes the speed of light as experienced by the electrons under electromagnetic interactions, differing from the vacuum value of $c$ (values are discussed below). We solve for $\mathcal{H_{AC}}\Psi(x,y)=E\Psi(x,y)$, with 
\begin{equation}
\mathcal{H_{AC}} = \frac{1}{2m}(\bm{p}-\frac{1}{c^2}\bm{\mu}\times\bm{\mathcal{E}})^2+\frac{1}{2}m\,\omega_p^2\,y^2
\label{hacfull}
\end{equation}
Whereas the Aharonov-Bohm (AB) effect \cite{aharonov1959} uses the line integral $\oint q\bm{A}\cdot\bm{dl}$ to describe the phase accumulated by a charged particle along a trajectory that encloses a magnetic field flux, its electromagnetic dual, the AC effect \cite{aharonov1984} uses $(1/c^2)\oint(\bm{\mu}\times\bm{\mathcal{E}})\cdot\bm{dl}$ for the phase accumulated by a magnetic moment along a trajectory in an electric field. The duality \cite{hagen1990} is schematically illustrated in the interferometric ring geometries of Fig.~\ref{figluher11}b, as would be used to detect the effects in mesoscopic experiments. The AC effect was experimentally confirmed using neutron beam interferometry \cite{cimmino1989}. In the solid-state its similarity to the materials property of spin-orbit interaction (SOI) has led to experiments \cite{nitta2006,koenig2006,lillianfeld2010} and theoretical work \cite{mathur1992,balatsky1993,wang2005}. The Landau-like levels induced by the AC vector potential for various profiles of $\bm{\mathcal{E}}$ were described in Ref. \cite{bruce2005}. 

\begin{figure}
\includegraphics[width=83mm]{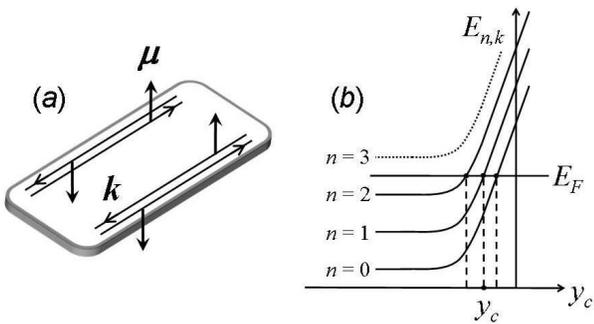}
\caption{\label{figluher21}(a): Schematic depiction of the helical edge states induced by the AC vector potential. If at a given edge the signs of $k$ and $\mu_z$ are changed simultaneously, the location of the edge state does not change. Counterpropagating states of opposite $\mu_z$ but same $n$ are in fact superposed. (b): Schematic depiction of $E_{n,k}$ \textit{vs} the wave function center point $y_c$. As in the IQHE, propagating states (labeled by $n$, $k$ and here also $\mu_z$) exist at $y$-coordinates where $E_F$ intersects $E_{n,k}$.}
\end{figure}

In Eq.~(\ref{hacfull}) and in Fig.~\ref{figluher11}a-b, we for now consider $\bm{\mu}\parallel\hat{z}$, such that $\bm{\mu}=(0,0,\mu_z)$. We will be able relax this requirement later, since only the product $\bm{\mu}\times\bm{\mathcal{E}}$ is involved. We introduce $\omega_A=\mu_z\omega_p^2/(qc^2)=\frac{1}{2}\omega_p(\mu_z/\mu_B)(\hbar\omega_p/mc^2)$, with $\mu_B$ the Bohr magneton. Here $\omega_A$ holds the same role in $\mathcal{H_{AC}}$ as the cyclotron frequency $\omega_C=qB/m$ holds in the IQHE. Using a wave function localized in $y$ and with plane-wave nature along $x$ over a sample of length $L$, $\Psi(x,y)=\frac{1}{\sqrt{L}}\,e^{ikx}\,\zeta(y)$, we obtain for the transverse function $\zeta(y)$:
\begin{equation}
\big[\frac{1}{2m}(\hbar k-m\,\omega_A\,y)^2+\frac{p_y^2}{2m}+\frac{1}{2}m\,\omega_p^2\,y^2\big]\zeta(y)=E\,\zeta(y)
\label{haciny}
\end{equation}
which, using a procedure identical to that yielding edge states in the IQHE \cite{macdonald1984,zawadzki1989}, can be rewritten as a harmonic oscillator shifted in $y$: 
\begin{equation}
\big[\lambda\frac{\hbar^2 k^2}{2m}+\frac{p_y^2}{2m}+\frac{1}{2}m\Omega^2(y-\eta k)^2\big]\zeta(y)=E\,\zeta(y)
\label{hacosciny}
\end{equation}
with $\lambda=\omega_p^2/(\omega_p^2+\omega_A^2)$, $\Omega=\sqrt{\omega_p^2+\omega_A^2}$ and $\eta=\hbar\omega_A/[m(\omega_p^2+\omega_A^2)]$. We find that the energy, $E_{n,k}=\lambda(\hbar^2 k^2)/(2m)+\hbar\Omega(n+\frac{1}{2})$, and that $\zeta_{n,k}(y)=e^{-q^2/2}H_n(q)$ with $q=y-\eta k$ and with $H_n(q)$ the $n^{th}$ Hermite polynomial. The transverse wave function is hence centered at $y_c=\eta k$. A similar spatial separation was derived in the context of spin accumulation at the boundaries under confinement-induced SOI \cite {hattori2006,jiang2006,xing2006} (\textit{cfr} below) or in the context of quantized magnetization transport \cite {meier2003}. However, given the mapping of Eqs.~(\ref{hacfull})-~(\ref{hacosciny}) on IQHE equivalents, an analysis in terms of edge states is compelling. The velocity of the state is $\frac{1}{\hbar}\frac{dE_{n,k}}{dk}=\lambda(\frac{\hbar k}{m})$. The position $y_c=\eta k$ and the velocity are hence both proportional to $k$. Yet, $y_c$ also depends on $\mu_z$, via $\omega_A$. To conclude the same position $y_c$ and same energy $E_{n,k}$ for given states, the signs of $k$ and $\mu_z$ must be changed simultaneously. Hence at the same edge, states with $\mu_z>0$ propagate in a direction opposite to states with $\mu_z<0$, as illustrated in Fig.~\ref{figluher21}a. Figure~\ref{figluher21}b schematically depicts the energy dispersion close to the edge.  At the edge, the Fermi level $E_F$ cuts through a finite number of dispersion curves $E_{n,k}$, labeled by $n$, at their respective locations $y_c$, with higher $n$ corresponding to locations further removed from the edge. In the bulk the $E_{n,k}$ develop a gap, in analogy with Landau levels generating edge states in the IQHE. Equation~(\ref{hacfull}) applies for an arbitrary $\mu$, and in particular applies for the moment from a particle of spin 1/2. In that case, $\mu_z>0$ and $\mu_z<0$ are naturally identified with the two projections of the spin along the quantization axis \cite{wang2005,kohda2008}. In the case of spin 1/2, after solving Eq.~(\ref{hacfull}), an observer will hence conclude to the existence of helical edge states, as depicted in Fig.~\ref{figluher21}a. Helical edge states (based on the projection of spin in the $z$-direction) are also found in the QSHE \cite{kane2005b,koenig2007}. Backscattering at one edge in Fig.~\ref{figluher21}a requires $\bm{\mu}$ to be flipped, $\mu_z\rightarrow -\mu_z$, an operation which, involving a magnetic quantity, requires time reversal symmetry to be broken. Hence, unless scattering potentials are present whereby time reversal symmetry is broken, an incident electron will be transmitted across disordered regions, and backscattering is suppressed as the temperature $T\rightarrow 0$. Thus the physical picture closely parallels the QSHE, and a similar reasoning emerges regarding measurable quantities \cite{kane2005b}. The helical edge states here arise in a simple mesoscopic wire with parabolic confinement potential, under the action of the AC vector potential, and do not require the special band structure under which the QSHE was so far described \cite{kane2005b,bernevig2006b,koenig2007}. 

The electric field $\bm{\mathcal{E}}$ assumes two roles: the linearity of $\mathcal{E}$ in $y$ over the width of the sample leads to an AC vector potential equivalent to the magnetic vector potential of a homogeneous $\bm{B}$, and, the resulting parabolic $V(x,y)$ defines the sample's edges. The double role is illustrated by $\Omega=\sqrt{\omega_p^2+\omega_A^2}$, where both $\omega_A$ and $\omega_p$ depend on the existence of the parabolic potential, whereas $\omega_C$ is independently created by $\bm{B}$ in the IQHE equivalent $\Omega=\sqrt{\omega_p^2+\omega_C^2}$. The need in Eq.~(\ref{hacfull}) to maintain $\bm{\mathcal{E}}$ across the sample width indicates a particular relevance for narrow channels, such as encountered in mesoscopic experiments and point-contacts. The energy gap $\hbar\Omega$ yields an insulating bulk when $E_F$ lies within this gap, while the edges remain metallic (Fig.~\ref{figluher21}b). The gap isolates the edge states and plays an important role in the IQHE and QSHE. We should thus ascertain how deviations from parabolicity in $V(x,y)$ affect the isolation of the helical edge states. To first order, we find that a perturbation symmetric in $y$ alters the magnitude of the gap but does not qualitatively affect the isolation of edge states. A perturbation asymmetric in $y$ leads to a shift in $E_{n,k}(y_c)$ and can affect the isolation if of sufficient strength, establishing the benefits of a symmetric $V(x,y)$. With a vector potential at hand in Eq.~(\ref{hacfull}), we can now follow the Laughlin gauge-invariance argument for quantization in the IQHE \cite{laughlin1981,halperin1982}. We find that the AC flux through Laughlin's cylinder is now quantized as  $(1/c^2)\oint\bm{\mu}\cdot(\bm{\mathcal{E}}\times\bm{dl})=n\,2\pi\hbar$. In the dual effect magnetization, rather than charge, at the sample edges is expected to be quantized, as is indeed predicted for the QSHE \cite{kane2005b}. In the broadest terms, we recognize that the AC vector potential, when transformed from the magnetic vector potential, introduces phenomena at the edge of a narrow channel.  

We now compare the results from Eq.~(\ref{hacfull}) to confinement-induced SOI. The SOI term in the Hamiltonian can be written as $\mathcal{H_{SO}}=\beta\bm{\sigma}\cdot(\bm{k}\times\bm{\mathcal{E}})$, where $\bm{\sigma}$ is the vector of Pauli matrices, and where $\bm{\mathcal{E}}$ is identified with the in-plane confinement electric field \cite {hattori2006,jiang2006,debray2009}. With the $y$-component of $\bm{\mathcal{E}}$ as $\mathcal{E}=-(m/q)\,\omega_p^2\,y$, with $\bm{k}\parallel\hat{x}$ and considering the projection of spin along $\hat{z}$, we can write $\mathcal{H_{SO}}=-\hbar k\,\omega_S\,y$. This defines $\omega_S$, with a role equivalent to $\omega_A$ above. In analogy to Eq.~(\ref{haciny}), we find that the transverse function $\zeta(y)$ follows: 
\begin{equation}
\big[\frac{\hbar^2 k^2}{2m}-\hbar k\,\omega_S\,y+\frac{p_y^2}{2m}+\frac{1}{2}m\,\omega_p^2\,y^2\big]\zeta(y)=E\,\zeta(y)
\label{hsoi}
\end{equation}
We rescale the potential term in Eq.~(\ref{hsoi}) to a weaker confinement potential $\frac{1}{2}m\,\omega_p^{'2}\,y^2$, with $\omega_p^{'2}=\omega_p^2-\omega_S^2$, and recover a form identical to Eq.~(\ref{haciny}) with the substitution of $\omega_p\rightarrow \omega_p^{'}$ (rescaling of the potential does not affect the physical conclusions). After rewriting Eq.~(\ref{hsoi}) as a shifted harmonic oscillator, the energy is expressed as $E_{n,k}=\lambda'(\hbar^2 k^2)/(2m)+\hbar\Omega'(n+\frac{1}{2})$, with $\lambda'=\omega_p^{'2}/(\omega_p^{'2}+\omega_S^2)$ and $\Omega'=\sqrt{\omega_p^{'2}+\omega_S^2}$. The transverse wave function becomes $\zeta_{n,k}(y)=e^{-q^2/2}H_n(q)$ with $q=y-\eta' k$ where $\eta'=\hbar\omega_S/[m(\omega_p^{'2}+\omega_S^2)]$. Hence $\zeta_{n,k}(y)$ is now centered at $y_c=\eta' k$. Equation~(\ref{hsoi}), in complete analogy to Eq.~(\ref{haciny}), will then indeed yield spatially-separated helical edge states, counterpropagating for opposite spins. The helical edge states induced by $\mathcal{H_{SO}}$ are thus equivalent to the AC edge states. Such SOI states are of interest in mesoscopic geometries, particularly in split-gate point-contacts. A parabolic $V(x,y)$ with an in-plane $\bm{\mathcal{E}}$ however only approximates the complex three-dimensional $\bm{\mathcal{E}}$ encountered in split-gate point-contacts \cite{nixon1991,hsiao2009}. For instance, a $y$-dependent $z$-component of $\bm{\mathcal{E}}$, $\mathcal{E}_z(y)$, likely  exists in these systems, in addition to the in-plane component of $\bm{\mathcal{E}}$. As mentioned above however, only $\bm{\mu}\times\bm{\mathcal{E}}$ matters, and with Bychkov-Rashba SOI \cite{bychkov1984} aligning spin perpendicular to $\bm{k}$ and to $\hat{z}$, $\mathcal{E}_z(y)$ can then lead to a term analogous to $-\hbar k\,\omega_S\,y$. We conclude that in point-contacts, either the $y$- or inhomogeneous $z$-component of $\bm{\mathcal{E}}$ can yield helical edge states, with implications for transport phenomena \cite{jphysc2008}. 

We now outline experimental challenges to observe the helical states from Eq.~(\ref{hacfull}). We assume that the observer uses the electron spin for $\bm{\mu}$ and the confinement field for $\bm{\mathcal{E}}$. As described above for SOI, the term $\bm{\mu}\times\bm{\mathcal{E}}$ can in mesoscopic geometries either arise from the projections of spin along $\hat{z}$ (via $\mathcal{E}_y$) or along $\hat{y}$ (via $\mathcal{E}_z(y)$). To maximize $\mu$ and minimize the effective $c$, we consider the quasi-relativistic \cite{zawadzki1985} narrow-gap semiconductors with large electron g-factors ($g$), InGaAs \cite{kohda2008}, InAs \cite{brosig2000} or InSb \cite{kallaher2010a,kallaher2010b}. The band structure implies a momentum \textit{vs} energy response differing from that in vacuum, determining the electron dynamics under electromagnetic fields. Non-parabolicity in the conduction band can be expressed in terms of an effective $c\approx\sqrt{E_g/2m^*}$, where $E_g$ denotes the bandgap and $m^*$ the effective mass at the $\Gamma$-point \cite{zawadzki1985}. For InGaAs, InAs and InSb it is found that $c\approx\,1.2\times10^6 m/s$, about 250 times lower than the vacuum value. The magnetic moment is considered as $\mu=\frac{1}{2}g\,\mu_{B}$. As all three materials yield similar estimates, we consider a 2DS in InAs ($\Gamma$-point $m^*$=0.024 $m_e$ with $m_e$ the free electron mass, and $g$=-15). At a 2D density $10^{12}\,cm^{-2}$, taking non-parabolicity into account, $E_F=83\:meV$. To derive a value for $\omega_p$ we use $E_F$ as the approximate classical turning point of $V(x,y)$, and assume a depletion layer width of 0.15 $\mu m$ at the edge, within the range of values encountered in 2DS (depletion layers in InSb 2DSs \cite{hongchen2005,kallaher2010b} are wider than in InAs 2DSs \cite{lillianfeld2010}, likely due to the accumulation layer present at InAs surfaces \cite{tsui1975}). We find $\mathcal{E}\approx 1\times 10^6 V/m$ (similar values of $\mathcal{E}$, about an order of magnitude below breakdown, are typical in semiconductor heterostructures, justifying the approach). With these values we find $\hbar\omega_A\approx 0.01\: meV$, corresponding to 0.13 $K$. It is enlightening to cast this estimate in an IQHE equivalent: $\hbar\omega_A$ corresponds to $\sim 7\:mT$ in a GaAs 2DS IQHE experiment. While allowing the interferometric mesoscopic ring experiments mentioned above, such fields point to experimental challenges for experiments closely copying the standard IQHE geometries. Equation~(\ref{hacfull}) also predicts Shubnikov-De Haas-like oscillations in magnetotransport, from density-of-states effects. Such electrical measurements avoid difficulties encountered with measuring magnetization properties in the QSHE. In the edge state regime, one can also envision measurements based on altering the edge state structure via side gates, or using applied magnetic fields to align $\bm{\mu}$ parallel to $\bm{\mathcal{E}}$. 

In conclusion, we address Aharonov-Casher edge states in a narrow channel. Helical edge states are predicted, which closely share several features with the edge states in the recently-described QSHE. The vector potential transformation used in this work raises the possibility that other sets of closely-related effects may exist. We acknowledge illuminating discussions with S. Ren, D. Minic, K. Park and V. Scarola, and this work is supported by DOE through award DE-FG02-08ER46532.

\end{document}